\documentclass[useAMS,usenatbib]{mn2e}

\input{epsf}
\usepackage{graphicx}
\usepackage{longtable}
\usepackage{lscape}

\def\Hb{H${\beta}$}
\def\Hg{H${\gamma}$}
\def\O3{\mbox{[O\,{\sc iii}]}}

\def\aj{AJ}%
\def\araa{ARA\&A}%
\def\apj{ApJ}%
\def\apjl{ApJ}%
\def\apjs{ApJS}%
\def\aap{A\&A}%
\def\mnras{MNRAS}%

\title[Quasar Variability and Black Hole Mass]{The Dependence of
Quasar Variability on Black Hole Mass} \author[M. Wold,
M. S. Brotherton, and Z. Shang]{M. Wold$^{1}$\thanks{E-mail:
mwold@eso.org}, M. S. Brotherton$^{2}$\thanks{E-mail:
mbrother@uwyo.edu (MSB), ESO Visitor}, and Zhaohui
Shang$^{2}$\thanks{E-mail: shang@uwyo.edu}\\ $^{1}$European Southern
Observatory, Karl-Schwarzschildstr.\ 2, 85748 Garching bei M{\"u}nchen,
Germany\\ $^{2}$Department of Physics and Astronomy, University
of Wyoming, Laramie, WY, 82072, U.S.A.\\ }

\begin{document}


\pagerange{\pageref{firstpage}--\pageref{lastpage}} \pubyear{2006}

\maketitle

\label{firstpage}

\begin{abstract}

In order to investigate the dependence of quasar variability on
fundamental physical parameters like black hole mass, we have matched
quasars from the QUEST1 Variability Survey with broad-lined objects
from the Sloan Digital Sky Survey. The matched sample contains
$\approx100$ quasars, and the Sloan spectra are used to estimate black
hole masses and bolometric luminosities. Variability amplitudes are
measured from the QUEST1 light curves. We find that black hole mass
correlates with several measures of the variability amplitude at the
99\% significance level or better. The correlation does not appear to
be caused by obvious selection effects inherent to flux-limited quasar
samples, host galaxy contamination or other well-known correlations
between quasar variability and luminosity/redshift.  We evaluate
variability as a function of rest-frame time lag using structure
functions, and find further support for the variability--black hole
mass correlation.  The correlation is strongest for time lags of the
order a few months up to the QUEST1 maximum temporal resolution of
$\approx2$ years, and may provide important clues for understanding
the long-standing problem of the origin of quasar optical variability.
We discuss whether our result is a manifestation of a relation between
characteristic variability timescale and black hole mass, where the
variability timescale is typical for accretion disk thermal
timescales, but find little support for this. Our favoured
explanation is that more massive black holes have
larger variability amplitudes, and we highlight the need for larger samples
with more complete temporal sampling to test the robustness of this result.

\end{abstract}


\section{Introduction}

Quasars have been recognized as optically variable since their
discovery, varying on timescales from hours to decades. The optical
variability is, at best, characterized as poorly understood, but is
nevertheless recognized as a means of probing physical scales that
cannot be resolved spatially by any telescope or instrument
\citep[e.g.][]{bmk82,np97,peterson04}. Variability is therefore an
important diagnostic of the physical processes responsible for 
the activity of active galactic nuclei (AGN).

Quasar light curves do not show evidence for periodicities, eclipses,
or any other easily understood signature. However, optical-UV quasar
variability has been studied for quite some time, and observational
trends between variability and other quasar properties such as
luminosity and redshift are comfortably established, and selection
effects reasonably well understood (Giallongo, Trevese \& Vagnetti
1996; Cristiani et al. 1996 Cid Fernandes, Aretxaga \& Terlevich 1996;
Ulrich, Maraschi \& Urry 1997). For instance, variability is found to
be inversely correlated with the optical luminosity of the quasar, and
it has been demonstrated that, in their brighter phases of variability,
quasars become bluer
\citep[e.g.][]{as72,hook94,cf96,giveon99,wm00,vandenberk04}.
Such a bluening, or hardening, of quasar spectra during  
their brighter phases is likely due to the fact that two different
spectral components with different variability properties make up the
continuum \citep[e.g.][]{ulrich97}.  Quasars are also found to be
increasingly variable at longer timescales, at least up to timescales
of years \citep{hook94,cristiani96,cp01}. Finally, a correlation
between variability and redshift is also observed, with quasars being
more variable at higher redshifts \citep[e.g.][]{cristiani90,hook94}.
This correlation is apparently understood as a selection effect
arising because of the chromatic nature of the variability.  As
quasars are more variable in the blue, and because shorter rest-frame
wavelengths are probed at higher redshifts, the net effect is a
correlation between redshift and variability
\citep{giallongo91,cristiani96,cf96}.

Radio-selected quasars do not show the anti-correlation between
variability and optical luminosity, but are seen to be bluer 
in the brighter phases of their variability, just as optically selected
quasars. 
\citep{helfand01,enya02,vandenberk04}. Helfand et al.\ find that
radio-selected quasars also do not show a correlation between
variability and radio luminosity, but that the most radio-loud quasars
may be marginally more variable than the radio-quiet \citep[see
also][]{enya02,vandenberk04,rengstorf06}. Some radio-loud quasars, the blazars,
are extremely variable on short timescales and likely have
relativistically beamed jets pointed close to the line of sight.  This
kind of variability is understood at some level, and is less
mysterious than the more general fluctuations seen in unbeamed quasar
continua.

\citet{vandenberk04}, analyzing by far the largest sample of quasars
($>$ 25,000 from the Sloan Digital Sky Survey or SDSS) claim that
there is evidence for redshift evolution in quasar variability with
quasars becoming more variable at higher redshifts. They suggest that
this may reflect changes in the quasar population or in the mechanism
causing the variability.  Note that this effect is evolutionary and
different from the
variability--redshift correlation observed in other samples and
explained as a selection effect
\citep{giallongo91,cristiani96,cf96,tv02}. 

A number of models have been proposed to explain optical-UV quasar
variability and the observed trends with quasar properties. At the
most fundamental level, and clearly most important on the longest
timescales, is the accretion rate.  It is less clear if accretion rate
variations, and perhaps corresponding disk temperature changes, are at
the heart of variations on the timescales of years, although such
changes can also account for the color changes
\citep[e.g.][]{pereyra06}.  Disk instabilities have also been invoked
\citep{kawaguchi98}.  Others have proposed that stellar processes
contribute, such as stellar collisions \citep{Torricelli00} or
supernovae (Terlevich et al.\ 1992; Aretxaga \& Terlevich 1994;
Aretxaga, Cid Fernandes \& Terlevich 1997; Cid Fernandes, Aretxaga \&
Vieira da Silva 2000).  Microlensing is an important source of variability
in some lensed quasars \citep[e.g.][]{refsdal00,morgan06}, and may
more generally be important.  Processes intrinsic to the central
engine itself must dominate, at least above some luminosity level,
since photoionized emission lines are seen to respond to continuum
changes after some delay.

It has been difficult to distinguish between the different models from
existing observational data. The majority of the proposed models have
all been shown to be qualitatively in agreement with observations.
One way of attempting to narrow down the number of possible models,
and also to help constraining existing models, is to find
relationships between variability and other AGN parameters, such as
black hole mass.  The black hole mass is a fundamental parameter of
the AGN, and the discovery of such a relationship -- or lack thereof
-- may provide additional clues to the physical mechanisms behind the
variability.

Quasar variability has already led to the measurement of black hole
masses.  Reverberation mapping uses spectrophotometric monitoring to
determine the time lag between continuum variation and the response
from broad emission lines like H$\beta$.  Due to the finite speed of
light, this time lag corresponds to a size scale of the line-emitting
region.  In combination with the velocity dispersion of the gas in the
variable line-emitting region, the size scale can be used to infer a
virial mass for the central black hole \citep[e.g.][]{peterson04}.
The time lags, and sizes, correlate with quasar luminosity
\citep[e.g.][]{kaspi00,kaspi05}, and the instantaneous velocity
dispersion may therefore be used as a stand-in for the variable
component of the line width, thus permitting single-epoch observations
of quasars to be used for estimating black hole masses
\citep{vestergaard02,vp06}.

The capability of making single-epoch estimates of black hole masses,
even with uncertainties of factors of several, is powerful since AGN
black hole masses span orders of magnitudes, from millions to billions
of solar masses, and tens of thousands of spectra are available from
recent surveys.  Less readily available are good quality light curves
of very many quasars, although surveys taking advantage of new
technologies are making up for this.  We have utilized these new
surveys to search for relationships between quasar variability and
fundamental AGN parameters, like black hole mass.

In the next section we describe how we defined a sample of quasars by
matching sources from the QUEST1\footnote{QUasar Equatorial Survey
Team, Phase 1} Variability Survey \citep{rengstorf04b}
and the SDSS data release 2 (DR2) \citep{abazajian04}. We explain how
data from the two surveys were used to obtain measurements of
variability and black hole mass. The analysis is carried out as
described in Section~\ref{section:analysis} using correlation statistics
and structure functions. A number of selection effects are
investigated as possible causes for the observed correlation between
black hole mass and variability amplitude, and we discuss the results
in Section~\ref{section:discussion}. The conclusions are drawn in
Section~\ref{section:conclusions}.

Throughout we have assumed a cosmology with $H_{0}=70$
km\,s$^{-1}$\,Mpc$^{-1}$, $\Omega_{m}=0.3$ and $\Omega_{\Lambda}=0.7$.

\section{Sample, Data, and Measurements}
\label{section:sec2}

Table~\ref{table:t1} lists the 104 quasars in our sample, along with
black hole masses, bolometric luminosities,
Eddington ratios, and assorted variability parameters.  The following
sections provide details.

\subsection{Sample}

The sample was formed by matching objects categorized as having broad
emission lines at redshifts $z<0.75$ in the SDSS DR2
\citep{abazajian04} with sources in the 200k Light Curve Catalogue of
the QUEST1 Variability Survey \citep{rengstorf04b}. The
redshift constraint was chosen for two reasons. Firstly, to ensure
that the H$\beta$ would lie within the SDSS spectral coverage hence
permitting consistent black hole mass estimates using well established
techniques, and, secondly, to limit the extent of time dilation
effects which might introduce biases.

We find 108 matches meeting our selection criteria, with 86 sources
recovered in the SDSS QSO catalogue \citep[based on
DR3]{schneider05}. The 22 sources that were not recovered in the QSO
catalogue are among the less luminous and probably did not satisfy the
absolute magnitude criterion of the SDSS QSO catalogue.  We include
the 22 sources, which all have typical quasar spectra, in our sample.
Four sources were rejected on the basis of poor spectral quality,
hence our final sample consists of 104 quasars.

\subsection{Black Hole Mass Estimation}

We estimate virial black hole masses for the quasars based on
single-epoch SDSS spectra and the scaling relationships and formalism
of \citet{vp06}. In order to to this, a measurement of the velocity
dispersion of the H$\beta$ line and the continuum luminosity are
required.

\begin{figure}
\begin{center}
\includegraphics[width=8.4 truecm]{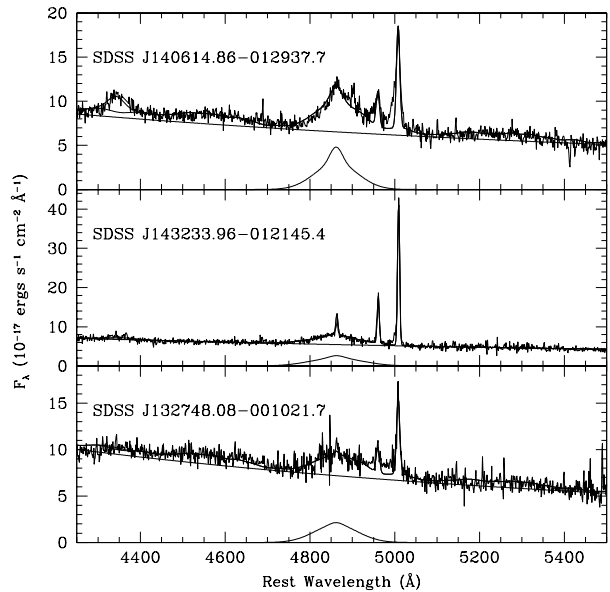}
\end{center}
\caption{Three examples of SDSS spectra and typical fits to the
H$\beta$ region used to make measurements of the line width and
continuum flux needed for black hole mass estimation. Total,
continuum, and line components are shown.}
\label{figure:fig1}
\end{figure}

We begin with the SDSS spectra, which were dereddened using the
Galactic extinction values of \citet*{sfd98}.  Thereafter, the H$\beta$
region of each spectrum was fitted using the same techniques as
\citet{shang05}.  The spectra are fitted from 4250 to 6000 \AA\, using
the IRAF task {\em specfit} \citep{kriss94} and the region is modeled
with the following components: a power-law continuum, a broadened
optical Fe\,{\sc ii} template \citep{bg92}, a host galaxy contribution
based on stellar synthesis population models \citep{bc03}, and the
\Hg, \Hb, and \O3 $\lambda\lambda$4959,5007 emission lines. 
The \Hb\, line is fit with two broader Gaussians, plus a very narrow Gaussian
constrained to match the width of the \O3 line to account for the
occasionally strong narrow-line region contribution to \Hb. A velocity
shift is allowed between the components to account for the asymmetry
of the line profile.  The two Gaussian components do not necessarily
have any physical significance, but are required to fit the line well
enough to determine a reliable FWHM.  The initial parameters are set
such that the power-law continuum dominates the emission, but the
Fe\,{\sc ii} and the host galaxy can contribute when
necessary. Fig.~\ref{figure:fig1} shows examples of our
fitting results. For details about the fitting procedure, see
\citet{shang05}.

\begin{figure}
\begin{center}
\includegraphics[width=7.0 truecm]{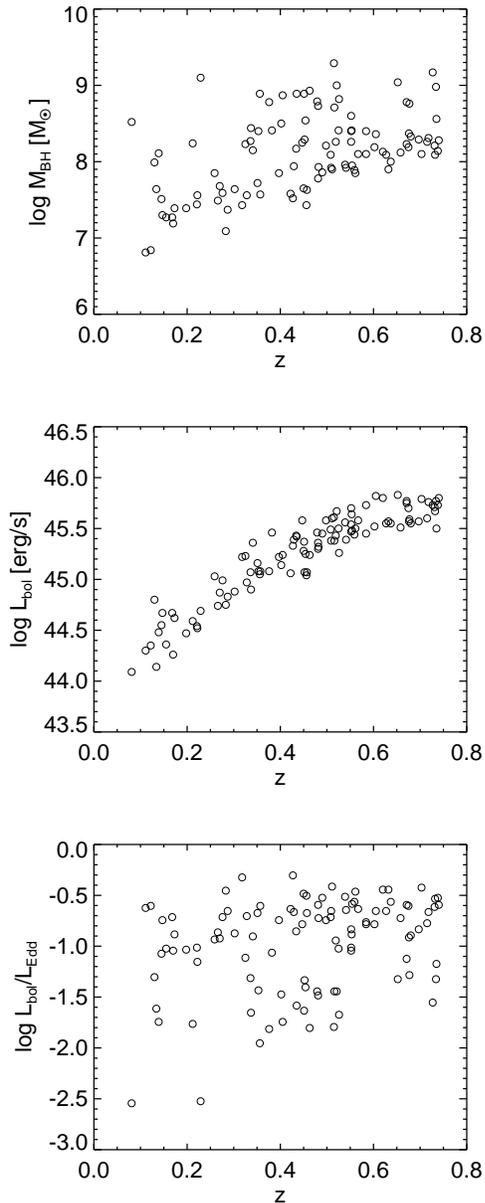}
\end{center}
\caption{Plots of three fundamental AGN properties against redshift
for the SDSS-QUEST1 sample.}
\label{figure:fig2}
\end{figure}

In order to estimate black hole masses, we utilize the FWHM of the
H$\beta$ line and the 5100 \AA\, continuum flux determined by the
fitting routine.  The FWHM of the H$\beta$ line is computed
numerically from the sum of the two Gaussian components used to fit
the line. The largest source of uncertainty related to determining the
FWHM is associated with the placement of the continuum, and we
estimate typical uncertainties of 10--20\%. The continuum luminosity
is obtained from the power-law component flux value at rest-frame 5100
{\AA}.

The bolometric luminosity is estimated as $L_{\rm bol} = 9 \times
\lambda L_{\lambda}5100$. The Eddington luminosity is calculated as
$L_{\rm Edd} = 1.51 \times 10^{38} M_{\rm BH}/M_{\odot}$ ergs s$^{-1}$
\citep{krolik98}\footnote{Krolik (1998) derives the Eddington luminosity
using a reduced particle mass, while other authours, e.g.\ \citet{peterson97},
use the proton mass. This changes the coefficient in the formula for 
Eddington luminosity from 1.51 to 1.26. Our analysis and conclusions do not
depend on this coefficient},
and the Eddington ratio as $\eta = L_{\rm bol}/L_{E\rm dd}$. Black
hole masses, bolometric luminosities and Eddington ratios are listed
in Table~\ref{table:t1} for every quasar in the sample, and plotted as
a function of redshift in Fig.~\ref{figure:fig2} to illustrate
the parameter space explored by the sample.

\subsection{Variability Data}

The QUEST1 Variability survey provides light curves for nearly 200,000
objects in $B$, $V$ and $R$ filters over a 2\fdg4 wide strip centered
at declination $-$1\degr\, and covering the range 10$^{\rm h}$ to
15$^{\rm h}$30$^{\rm min}$ in right ascension
\citep{rengstorf04b}. All sources in the 200k Light Curve Catalogue
overlap with sources in the SDSS DR2. The limiting magnitudes of the
200k Light Curve Catalogue are $B\approx19.6$, $V\approx19.8$ and
$R\approx20.8$, and only point sources are included (typical seeing is
2\farcs8).  The QUEST1 light curves are given in terms of Julian date,
instrumental magnitude and magnitude uncertainty in four filters
denoted $b$, $v$, $r_{1}$ and $r_{3}$ (lower case used for
instrumental magnitudes). In order to check for self-consistency, the
QUEST1 survey uses two $R$ filters, denoted $r_{1}$ and $r_{3}$, hence
a criterion for inclusion in the 200k catalogue is that the object has
a significant detection in both $R$ filters. Our matched sample
therefore contains 104 quasars detected in each of the two $R$
filters. In the $V$-filter we found 70 matches, whereas in the
$B$-filter, only 11. Since a sample of 11 is too small for the
purposes of this paper, we have ignored the $B$-band light curves.  We
have also ignored the $r_{1}$ measurements, and chosen $r_{3}$ for our
analysis since the results for the two $R$-filters are very similar.

\begin{figure}
\begin{center}
\includegraphics[width=8.4 truecm]{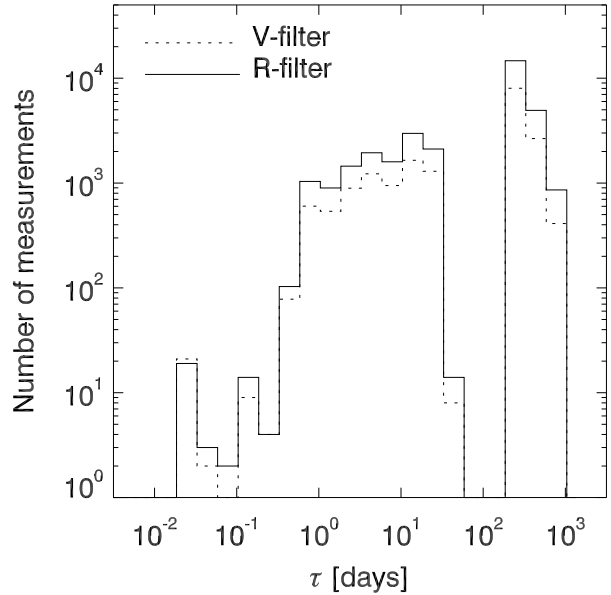}
\end{center}
\caption{The distribution of rest-frame time lags for the SDSS-QUEST1
sample.  The total number of objects in the $R$-filter is $N=104$, and
in $V$, $N=70$.}
\label{figure:fig3}
\end{figure}

The QUEST1 Variability Survey extended over a period of 26 months,
hence covers observed timescales ranging from a few hours up to 26
months. The time elapsed between two photometric measurements in the
rest-frame of the quasars, referred to as the ``time lag'', $\tau$,
scales with $(1+z)^{-1}$. The distribution of rest frame 
time lags for the 104 quasars in our matched sample is shown in
Fig.~\ref{figure:fig3}.  The majority of the measurements
are concentrated in two regions, one at $0.5 < \tau < 30$ days and the
other at $180 < \tau < 700$ days, with a gap around 100 days.  The
sample therefore probes variations on rest-frame timescales between
$\approx12$ hrs up to one month, and between six months up to almost
two years. The measurements extend all the way down to time lags of
$<1$ hr, but only for a few objects. 

\subsection{Measurements of Variability}
\label{section:variability_measurements}

\subsubsection{Statistical Measurements}

Variability can be characterized in several different ways, and we
have chosen quantities describing both the variability of each quasar
and the variability for the sample as a whole. Common for both
of these is that the distribution of all possible magnitude
differences (or variability amplitudes) on each light curve is
evaluated, i.e.\
\begin{equation}
\Delta m_{ij} = m_{i}-m_{j},
\end{equation} 
where $i<j$. A light curve with $N$ photometric measurements at
different time lags therefore has $N(N-1)/2$ different variability
amplitudes. The variability of a single quasar can be described by the
standard deviation, mean, median and maximum of its distribution of
variability amplitudes \citep[see e.g.][]{giveon99}.  We denote these
quantities $\sigma_{\Delta m}$, $\left<\Delta m\right>$, Med$(\Delta
m)$ and Max$(\Delta m)$, and list them in Table~\ref{table:t1} for
both the $V$- and the $R$-filter.  The standard deviation,
$\sigma_{\Delta m}$, was evaluated by weighting each magnitude with
the inverse of the uncertainty in the magnitude.

Also listed in Table~\ref{table:t1}, in the last column, is the
variability global confidence level, denoted GCL, from the QUEST1
Variability catalogue. It is defined by \citet{rengstorf04b} as a
weighted average of the variability confidence levels in every filter
the source was detected, and indicates the percentage probability that
a source is not variable purely by random fluctuations.  The mean and
median GCL of our matched sample is 79 and 88, respectively, and the
percentage of objects with GCL $>$ 85, which \citet{rengstorf04a}
choose for spectroscopic follow-up and classify as ``highly
variable'', is 58. Hence more than half of our sample contains highly
variable objects, reflecting the fact that most quasars are indeed
variable at some level.

\subsubsection{Structure Functions}
\label{section:sf}

\begin{figure}
\begin{center}
\includegraphics[width=8.4 truecm]{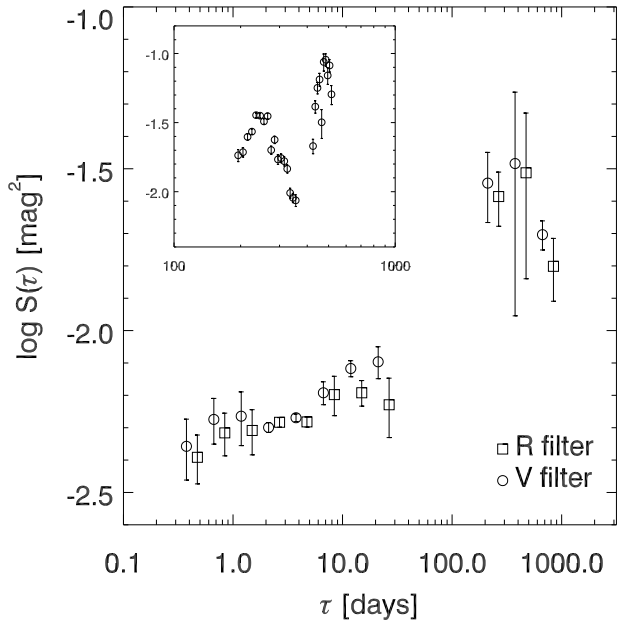}
\end{center}
\caption{Structure functions for the whole sample in both $R$- and
$V$-filter.  The smaller inlaid figure shows the structure function in
bins of 10 days at time lags between 100 and $\approx700$ days. }
\label{figure:fig4}
\end{figure}

The variability measurements outlined in the previous section
characterize how strongly variable each source is across the full
light curve, but do not describe how the variability changes as a
function of time lag. In order to investigate the time dependence of
the variability, we utilize the first-order structure function,
defined as
\begin{equation}
S\left(\tau\right) = \left< \left[m(t) - m(t+\tau)\right]^2 \right>
\end{equation}
\citep*{hughes92}, where $m(t)$ is the magnitude at time $t$, $\tau$ is
the rest-frame time lag, and the $\left< \right>$ brackets denote the
ensemble average. In practice, the structure function is the sample
average of all magnitude differences at a given time lag, $\tau$. It
is often used to characterize quasar variability
\citep{hughes92,vandenberk04,devries05} and is less sensitive to
aliasing problems and gaps in the data than e.g.\ Fourier analysis.
We evaluate the structure function for the sample within fixed
logarithmic intervals of width 0.25 in $\log \tau$. The resulting
structure function in both $V$- and $R$-filter is shown in
Fig.~\ref{figure:fig4} where only bins with more than 50
measurements have been included (the number of measurements per bin
can be read off Fig.~\ref{figure:fig3}).  The uncertainty
in $S(\tau)$ at each $\log \tau$ interval was evaluated by splitting
each bin in four equal parts and calculating the difference between
$S(\tau)$ in each sub-bin and $S(\tau)$ in the full bin.

The behaviour of the structure function in Fig.~\ref{figure:fig4} is
typical for measurements where there is variability due to measurement
noise at shorter time lags and variability intrinsic to the source
at longer time lags. For measurement noise, the correlation
timescale is zero, hence the plateau at short time lags, $\tau \la 5$
days, is characteristic of the photometric uncertainty in the data,
here seen to be $\approx0.05$--$0.06$ mag.  Since the plateau at short
time lags merely probes the noise, we expect the structure function to
be similar in both $V$ and $R$ in this region, which is indeed
observed. For time lags $\ga$ 6--7 days, the structure function starts
to increase. The increase is seen in both $V$ and $R$, but with the
sample being more variable in the $V$-filter.  This is in line with
previous observations, finding that quasars are more variable in the
blue.

At time lags between one and six months there are no data, but at
$\tau \ga 6$ months, the structure function has increased
significantly above the noise plateau. A structure function that
increases from the noise plateau and levels off at longer time lags is
typical of a process which has one characteristic, physically
meaningful, timescale \citep{hughes92}. Whether there is such a
plateau at longer timescales in our sample is difficult to determine
because there are not enough measurements at long time lags (only
three data points at $\log \tau > 2$). A leveling off at $\tau \approx
200$ days would indicate that there is a preferred timescale of
roughly 200 days for the outbursts causing the variability, whereas a
continued increase indicates that processes of many different
timescales contribute to the variability.

It is uncertain whether the turn-over of the structure function at 
$\tau \approx 300$--400 days reflects a real change. Most likely it 
is caused by incompleteness or selection effects.  Each bin, apart 
from the first and the last, contains
measurements from all objects in the sample. The bin at the longest
time lag contains objects that are biased toward smaller redshifts,
because only for those can we probe the longest time lags.  The
smaller plot in the upper left-hand corner of Fig.~\ref{figure:fig4}
shows the structure function evaluated in bins of 10 days.  There is a
dip around $\tau \approx 400$ days, causing the larger error bar in
the more coarsely binned data. The dip at $\approx400$ days, as well
as the decrease in $S(\tau)$ at longer time lags may therefore be due
to a bias introduced by incomplete sampling of time lags at the end of
the one- and two-year periods in the QUEST1 variability survey. As the
longest time sampling is obtained for the lower-redshift sources, and
if the lower-redshift sources are biased toward lower variability,
this could explain the drop in the structure function.
It is well known that quasar variability continues to increase
at time lags of the order several years 
\citep{hook94,vandenberk04,devries05}, hence 
the most likely explanation for the turn-over 
in this set of data is insufficient sampling at long time lags.
A similar turn-over is also seen in the structure function
analysis made by \citet{rengstorf06} on a sample of $\sim1000$ 
QUEST1 quasars.

Regardless of the behaviour of the structure function in the longer
time lag bin, 
the slope at the rising part is a meaningful parameter as it
indicates the nature of the process causing the variability
\citep{hughes92}. But as there is a gap in the data at time lags from
one to six months, we can only estimate a lower limit to the
slope. Fitting a straight line through the data points at $\tau > 6$
days gives a slope of 0.42 in $V$ and 0.41 in $R$. 
This agrees very well with the slope of 0.41$\pm$0.07 derived by 
\citet{rengstorf06} for their sample of roughly 1000 QUEST1 quasars.

\begin{landscape}
\begin{table}
\begin{minipage}{20cm}
\caption{The total sample of 104 quasars with variability measurements
and AGN parameters. The columns are (1) Sloan ID number, (2) redshift,
(3) apparent Sloan $r$ magnitude, (4) absolute magnitude in Sloan
$i$-filter, (5)--(8) weighted standard deviation, mean, median and
maximum of the distribution of variability amplitudes in the
$R$-filter (units of magnitudes), (9)--(12) same as columns (5)--(8),
but for the $V$-filter, (13) log of black hole mass in units of solar
masses, (14) log of bolometric luminosity in units of ergs\,s$^{-1}$,
(15) log of Eddington ratio, defined as $L_{\rm bol}/L_{\rm Edd}$, and
(16) the global confidence level of variability as defined by
\citet{rengstorf04b}.}
\label{table:t1}
\begin{tabular}{lllllllllllllllr}
\hline
SloanID (SDSS J) & $z$ & $m_{r}$ & $M_{i}$ & $\sigma_{\Delta R}$ & $\left<\Delta R\right>$ & Med$(\Delta R)$ & Max$(\Delta R)$ & $\sigma_{\Delta V}$ & $\left<\Delta V\right>$ & Med$(\Delta V)$ & Max$(\Delta V)$ & $\log M_{\rm BH}$ & $\log L_{\rm bol}$ & $\log \eta$ & GCL \\
 (1) & (2) & (3) & (4) & (5) & (6) & (7) & (8) & (9) & (10) & (11) & (12) & (13) & (14) & (15) & (16) \\ 
\hline
100110.52$-$004049.1 & 0.134 & 18.59 & $-$20.31 & 0.060 & 0.073 & 0.057 & 0.310 & 0.063 & 0.071 & 0.070 & 0.207 & 7.640 & 44.14 & $-$1.61 & 75.42\\
100215.83$-$001056.1 & 0.353 & 18.80 & $-$22.51 & 0.083 & 0.098 & 0.087 & 0.295 & 0.101 & 0.115 & 0.105 & 0.358 & 8.400 & 45.08 & $-$1.43 & 71.77 \\
101003.14$-$001332.1 & 0.732 & 18.58 & $-$24.42 & 0.079 & 0.104 & 0.089 & 0.420 & 0.083 & 0.107 & 0.095 & 0.347 & 8.090 & 45.67 & $-$0.53 & 66.77 \\
101658.66$-$000708.4 & 0.337 & 18.92 & $-$22.34 & 0.301 & 0.262 & 0.156 & 0.809 & 0.000 & 0.000 & 0.000 & 0.000 & 8.440 & 44.90 & $-$1.65 &100.00\\
102504.34$-$004618.9 & 0.552 & 18.66 & $-$23.96 & 0.103 & 0.131 & 0.079 & 0.358 & 0.125 & 0.147 & 0.099 & 0.436 & 8.260 & 45.54 & $-$0.83 & 94.16 \\
102522.66$-$001552.4 & 0.452 & 19.57 & $-$22.71 & 0.091 & 0.102 & 0.086 & 0.361 & 0.000 & 0.000 & 0.000 & 0.000 & 8.290 & 45.07 & $-$1.33 & 79.43 \\
102606.73$-$005038.9 & 0.283 & 18.79 & $-$22.03 & 0.155 & 0.191 & 0.199 & 0.502 & 0.000 & 0.000 & 0.000 & 0.000 & 7.090 & 44.75 & $-$0.45 &100.00\\
102920.70$-$004747.6 & 0.259 & 18.29 & $-$22.45 & 0.040 & 0.050 & 0.046 & 0.141 & 0.070 & 0.088 & 0.069 & 0.291 & 7.850 & 45.03 & $-$0.93 & 98.36 \\
102926.13$-$002643.3 & 0.198 & 18.91 & $-$21.51 & 0.062 & 0.079 & 0.075 & 0.250 & 0.000 & 0.000 & 0.000 & 0.000 & 7.390 & 44.47 & $-$1.03 & 34.60\\
103031.41$-$001902.7 & 0.561 & 19.21 & $-$23.38 & 0.159 & 0.180 & 0.170 & 0.461 & 0.181 & 0.213 & 0.189 & 0.588 & 7.850 & 45.50 & $-$0.46 & 80.74 \\
103222.57$-$000345.6 & 0.559 & 19.31 & $-$23.25 & 0.103 & 0.123 & 0.102 & 0.386 & 0.075 & 0.087 & 0.079 & 0.281 & 7.890 & 45.44 & $-$0.56 & 69.38\\
103703.10$-$001854.9 & 0.287 & 19.13 & $-$21.98 & 0.069 & 0.085 & 0.083 & 0.253 & 0.000 & 0.000 & 0.000 & 0.000 & 7.370 & 44.83 & $-$0.65 & 39.01\\
104122.84$-$005618.4 & 0.498 & 18.54 & $-$23.69 & 0.043 & 0.051 & 0.043 & 0.167 & 0.033 & 0.039 & 0.034 & 0.098 & 8.210 & 45.58 & $-$0.74 & 71.06\\
104733.39$-$004700.5 & 0.740 & 18.46 & $-$24.44 & 0.041 & 0.049 & 0.044 & 0.157 & 0.046 & 0.055 & 0.047 & 0.208 & 8.280 & 45.80 & $-$0.59 & 48.56\\
105336.71$-$001727.3 & 0.482 & 19.19 & $-$23.00 & 0.056 & 0.069 & 0.062 & 0.216 & 0.068 & 0.078 & 0.069 & 0.241 & 7.930 & 45.32 & $-$0.72 & 40.50\\
105342.21$-$001420.1 & 0.676 & 19.13 & $-$23.76 & 0.060 & 0.068 & 0.061 & 0.206 & 0.070 & 0.072 & 0.063 & 0.245 & 8.370 & 45.57 & $-$0.91 & 37.58\\
105606.93$-$004655.5 & 0.336 & 19.06 & $-$22.24 & 0.061 & 0.074 & 0.058 & 0.337 & 0.069 & 0.077 & 0.064 & 0.277 & 8.270 & 45.07 & $-$1.31 & 52.13\\
105932.52$-$004354.8 & 0.155 & 18.76 & $-$20.91 & 0.056 & 0.065 & 0.057 & 0.199 & 0.058 & 0.073 & 0.067 & 0.244 & 7.270 & 44.36 & $-$1.02 & 30.09\\
110054.14$-$005038.4 & 0.541 & 19.39 & $-$23.22 & 0.064 & 0.080 & 0.072 & 0.220 & 0.000 & 0.000 & 0.000 & 0.000 & 7.920 & 45.39 & $-$0.64 & 36.58\\
111456.02$-$004019.2 & 0.734 & 19.01 & $-$24.04 & 0.095 & 0.119 & 0.121 & 0.371 & 0.081 & 0.106 & 0.090 & 0.367 & 8.980 & 45.77 & $-$1.32 & 96.39\\
111756.85$-$000220.5 & 0.457 & 19.24 & $-$22.78 & 0.065 & 0.080 & 0.072 & 0.236 & 0.000 & 0.000 & 0.000 & 0.000 & 7.630 & 45.07 & $-$0.67 & 48.17\\
112646.43$-$013417.9 & 0.341 & 18.55 & $-$22.74 & 0.052 & 0.062 & 0.055 & 0.227 & 0.086 & 0.096 & 0.087 & 0.307 & 8.150 & 45.36 & $-$0.90 & 87.12\\
112747.54$-$015830.1 & 0.519 & 18.93 & $-$23.48 & 0.064 & 0.081 & 0.071 & 0.232 & 0.084 & 0.100 & 0.091 & 0.315 & 8.260 & 45.43 & $-$0.94 & 33.88\\
113140.63$-$015118.3 & 0.434 & 19.13 & $-$22.88 & 0.231 & 0.235 & 0.121 & 0.611 & 0.000 & 0.000 & 0.000 & 0.000 & 8.170 & 45.43 & $-$0.85 &100.00\\
113318.89$-$004112.0 & 0.511 & 18.66 & $-$23.80 & 0.052 & 0.064 & 0.058 & 0.196 & 0.045 & 0.052 & 0.047 & 0.140 & 7.900 & 45.60 & $-$0.41 & 46.10\\
113416.81$-$001902.4 & 0.356 & 18.95 & $-$22.60 & 0.118 & 0.142 & 0.126 & 0.370 & 0.000 & 0.000 & 0.000 & 0.000 & 8.890 & 45.05 & $-$1.95 &100.00\\
113607.51$-$012313.7 & 0.266 & 18.96 & $-$21.84 & 0.093 & 0.102 & 0.086 & 0.384 & 0.000 & 0.000 & 0.000 & 0.000 & 7.490 & 44.74 & $-$0.86 & 98.18\\
114455.78$-$002142.7 & 0.351 & 18.52 & $-$22.85 & 0.152 & 0.160 & 0.128 & 0.455 & 0.127 & 0.151 & 0.139 & 0.446 & 7.720 & 45.16 & $-$0.67 &100.00\\
114528.56$-$004739.0 & 0.715 & 19.01 & $-$23.99 & 0.065 & 0.085 & 0.072 & 0.261 & 0.000 & 0.000 & 0.000 & 0.000 & 8.260 & 45.60 & $-$0.77 & 63.05\\
114718.05$-$013206.9 & 0.382 & 18.11 & $-$23.41 & 0.048 & 0.059 & 0.050 & 0.225 & 0.066 & 0.078 & 0.067 & 0.239 & 8.410 & 45.46 & $-$1.06 & 86.87\\
114758.38$-$001551.8 & 0.718 & 18.95 & $-$24.12 & 0.085 & 0.114 & 0.094 & 0.375 & 0.068 & 0.083 & 0.073 & 0.261 & 8.310 & 45.76 & $-$0.66 & 44.69\\
114916.75$-$004231.8 & 0.735 & 18.82 & $-$24.15 & 0.132 & 0.186 & 0.162 & 0.582 & 0.103 & 0.127 & 0.142 & 0.331 & 8.560 & 45.50 & $-$1.17 & 97.57\\
115156.69$-$011800.6 & 0.170 & 18.89 & $-$21.01 & 0.068 & 0.079 & 0.070 & 0.256 & 0.000 & 0.000 & 0.000 & 0.000 & 7.190 & 44.26 & $-$1.04 & 33.67\\
115216.12$-$005352.1 & 0.637 & 18.85 & $-$24.00 & 0.051 & 0.064 & 0.050 & 0.288 & 0.034 & 0.044 & 0.040 & 0.131 & 8.000 & 45.55 & $-$0.56 & 29.37\\
115306.94$-$004512.7 & 0.357 & 19.04 & $-$22.28 & 0.066 & 0.076 & 0.068 & 0.228 & 0.063 & 0.071 & 0.062 & 0.186 & 7.570 & 45.08 & $-$0.60 & 52.34\\
115342.99$-$001159.8 & 0.602 & 18.74 & $-$23.94 & 0.082 & 0.097 & 0.085 & 0.302 & 0.091 & 0.107 & 0.100 & 0.305 & 8.190 & 45.52 & $-$0.78 & 94.59\\
120553.64$-$004651.0 & 0.671 & 18.47 & $-$24.48 & 0.050 & 0.058 & 0.049 & 0.202 & 0.059 & 0.072 & 0.084 & 0.189 & 8.230 & 45.75 & $-$0.59 & 87.97\\
120619.01$-$003959.5 & 0.675 & 18.59 & $-$24.14 & 0.087 & 0.102 & 0.095 & 0.279 & 0.060 & 0.070 & 0.061 & 0.207 & 8.190 & 45.70 & $-$0.60 & 82.36\\
120629.59$-$004831.2 & 0.463 & 19.30 & $-$23.04 & 0.111 & 0.127 & 0.109 & 0.453 & 0.131 & 0.159 & 0.138 & 0.421 & 8.930 & 45.24 & $-$1.80 & 99.95\\
120644.82$-$012737.1 & 0.509 & 18.79 & $-$23.56 & 0.098 & 0.122 & 0.104 & 0.339 & 0.127 & 0.150 & 0.131 & 0.383 & 7.920 & 45.38 & $-$0.65 &100.00\\
\hline
\end{tabular} \\
\end{minipage}
\end{table}
\end{landscape}

\newpage

\begin{landscape}
\begin{table}
\begin{minipage}{20cm}
\contcaption{}
\begin{tabular}{lllllllllllllllr}
\hline
SloanID (SDSS J) & $z$ & $m_{r}$ & $M_{i}$ & $\sigma_{\Delta R}$ & $\left<\Delta R\right>$ & Med$(\Delta R)$ & Max$(\Delta R)$ &
$\sigma_{\Delta V}$ & $\left<\Delta V\right>$ & Med$(\Delta V)$ & Max$(\Delta V)$ & 
$\log M_{\rm BH}$ & $\log L_{\rm bol}$ & $\log \eta$ & GCL \\
 (1) & (2) & (3) & (4) & (5) & (6) & (7) & (8) & (9) & (10) & (11) & (12) & (13) & (14) & (15) & (16) \\ 
\hline
120708.57$-$013614.0 & 0.620 & 18.11 & $-$24.72 & 0.082 & 0.099 & 0.085 & 0.273 & 0.090 & 0.106 & 0.097 & 0.300 & 8.130 & 45.80 & $-$0.44 & 78.80\\
120806.89$-$013509.7 & 0.481 & 18.89 & $-$23.20 & 0.051 & 0.061 & 0.052 & 0.191 & 0.069 & 0.084 & 0.075 & 0.274 & 7.780 & 45.30 & $-$0.59 & 49.12\\
121224.35$-$015246.3 & 0.429 & 18.84 & $-$23.14 & 0.075 & 0.090 & 0.078 & 0.293 & 0.095 & 0.098 & 0.075 & 0.331 & 7.940 & 45.39 & $-$0.66 & 79.92\\
121337.06$-$000047.1 & 0.454 & 18.85 & $-$23.25 & 0.092 & 0.117 & 0.104 & 0.451 & 0.098 & 0.118 & 0.102 & 0.379 & 8.540 & 45.25 & $-$1.40 & 87.49\\
121449.89$-$011245.0 & 0.671 & 18.32 & $-$24.67 & 0.039 & 0.046 & 0.038 & 0.140 & 0.046 & 0.055 & 0.049 & 0.194 & 8.780 & 45.77 & $-$1.12 & 45.01\\
121747.38$-$015048.6 & 0.652 & 18.77 & $-$24.12 & 0.169 & 0.195 & 0.191 & 0.560 & 0.188 & 0.208 & 0.181 & 0.521 & 9.040 & 45.83 & $-$1.32 &100.00\\
122015.70$-$000251.6 & 0.397 & 19.05 & $-$22.68 & 0.101 & 0.118 & 0.114 & 0.362 & 0.097 & 0.116 & 0.101 & 0.318 & 7.850 & 45.22 & $-$0.74 & 92.07\\
122032.96$-$001434.1 & 0.212 & 18.81 & $-$21.75 & 0.075 & 0.081 & 0.069 & 0.306 & 0.000 & 0.000 & 0.000 & 0.000 & 8.240 & 44.59 & $-$1.76 & 84.62\\
122617.38$-$015955.3 & 0.584 & 18.90 & $-$23.79 & 0.095 & 0.119 & 0.101 & 0.414 & 0.000 & 0.000 & 0.000 & 0.000 & 8.100 & 45.45 & $-$0.76 & 99.18\\
123246.62$-$013639.9 & 0.325 & 17.75 & $-$23.11 & 0.053 & 0.065 & 0.054 & 0.181 & 0.115 & 0.134 & 0.122 & 0.358 & 8.230 & 45.23 & $-$1.11 &100.00\\
122347.96$-$013303.9 & 0.508 & 19.11 & $-$23.34 & 0.105 & 0.129 & 0.114 & 0.371 & 0.112 & 0.136 & 0.137 & 0.410 & 8.090 & 45.49 & $-$0.71 &100.00\\
122727.47$-$012158.2 & 0.456 & 19.37 & $-$22.86 & 0.105 & 0.126 & 0.103 & 0.492 & 0.133 & 0.169 & 0.150 & 0.511 & 7.430 & 45.04 & $-$0.50 & 99.74\\
123452.49$-$015955.6 & 0.328 & 18.78 & $-$22.30 & 0.108 & 0.125 & 0.112 & 0.392 & 0.000 & 0.000 & 0.000 & 0.000 & 7.560 & 44.97 & $-$0.70 &100.00\\
124519.73$-$005230.4 & 0.221 & 18.88 & $-$21.74 & 0.038 & 0.049 & 0.030 & 0.188 & 0.000 & 0.000 & 0.000 & 0.000 & 7.440 & 44.54 & $-$1.01 & 14.76\\
125055.28$-$015556.8 & 0.081 & 17.83 & $-$20.43 & 0.031 & 0.038 & 0.033 & 0.131 & 0.050 & 0.059 & 0.051 & 0.222 & 8.520 & 44.09 & $-$2.54 & 57.53\\
125337.35$-$004809.6 & 0.427 & 18.47 & $-$23.27 & 0.044 & 0.055 & 0.047 & 0.210 & 0.057 & 0.066 & 0.058 & 0.241 & 7.520 & 45.33 & $-$0.30 & 66.08\\
125952.49$-$015707.2 & 0.447 & 18.17 & $-$23.86 & 0.070 & 0.082 & 0.054 & 0.238 & 0.071 & 0.084 & 0.058 & 0.289 & 8.250 & 45.58 & $-$0.78 & 99.79\\
130023.21$-$005429.7 & 0.122 & 17.62 & $-$21.59 & 0.062 & 0.067 & 0.051 & 0.218 & 0.062 & 0.072 & 0.046 & 0.250 & 6.840 & 44.35 & $-$0.60 &100.00\\
130610.05$-$011600.6 & 0.229 & 18.47 & $-$22.04 & 0.108 & 0.133 & 0.116 & 0.319 & 0.000 & 0.000 & 0.000 & 0.000 & 9.100 & 44.69 & $-$2.52 &100.00\\
130725.69$-$004525.7 & 0.490 & 18.79 & $-$23.49 & 0.054 & 0.076 & 0.067 & 0.258 & 0.052 & 0.058 & 0.046 & 0.176 & 7.860 & 45.45 & $-$0.52 & 40.16\\
130707.70$-$002542.8 & 0.450 & 18.85 & $-$23.24 & 0.086 & 0.110 & 0.096 & 0.328 & 0.116 & 0.142 & 0.118 & 0.471 & 7.650 & 45.28 & $-$0.48 & 97.41\\
130845.68$-$013053.9 & 0.111 & 17.65 & $-$21.35 & 0.054 & 0.066 & 0.055 & 0.223 & 0.069 & 0.081 & 0.075 & 0.249 & 6.810 & 44.30 & $-$0.62 &100.00\\
130916.67$-$001550.1 & 0.422 & 19.15 & $-$22.66 & 0.095 & 0.123 & 0.106 & 0.370 & 0.000 & 0.000 & 0.000 & 0.000 & 7.580 & 45.06 & $-$0.63 & 97.18\\
130937.34$-$014950.4 & 0.703 & 18.42 & $-$24.56 & 0.089 & 0.103 & 0.093 & 0.360 & 0.098 & 0.127 & 0.115 & 0.423 & 8.100 & 45.79 & $-$0.42 &100.00\\
132231.12$-$001124.6 & 0.173 & 18.20 & $-$21.77 & 0.058 & 0.065 & 0.062 & 0.224 & 0.074 & 0.090 & 0.075 & 0.360 & 7.390 & 44.62 & $-$0.88 & 94.38\\
132514.79$-$012132.3 & 0.553 & 18.65 & $-$24.04 & 0.179 & 0.207 & 0.189 & 0.514 & 0.000 & 0.000 & 0.000 & 0.000 & 8.410 & 45.64 & $-$0.88 &100.00\\
132704.54$-$003627.5 & 0.302 & 19.10 & $-$21.95 & 0.070 & 0.087 & 0.068 & 0.376 & 0.000 & 0.000 & 0.000 & 0.000 & 7.640 & 44.88 & $-$0.87 & 71.27\\
132705.88$-$012415.5 & 0.168 & 17.71 & $-$22.21 & 0.048 & 0.057 & 0.041 & 0.269 & 0.064 & 0.079 & 0.069 & 0.244 & 7.270 & 44.67 & $-$0.71 & 99.95\\
132748.08$-$001021.7 & 0.479 & 18.97 & $-$23.30 & 0.125 & 0.140 & 0.110 & 0.494 & 0.000 & 0.000 & 0.000 & 0.000 & 8.790 & 45.46 & $-$1.44 &100.00\\
133105.30$-$005731.8 & 0.526 & 19.30 & $-$23.21 & 0.060 & 0.077 & 0.065 & 0.280 & 0.000 & 0.000 & 0.000 & 0.000 & 8.820 & 45.26 & $-$1.67 & 49.11\\
133141.02$-$015212.4 & 0.145 & 18.37 & $-$21.26 & 0.088 & 0.101 & 0.089 & 0.285 & 0.103 & 0.114 & 0.082 & 0.257 & 7.510 & 44.55 & $-$1.07 & 98.12\\
133350.26$-$003946.9 & 0.727 & 18.67 & $-$24.36 & 0.047 & 0.053 & 0.044 & 0.210 & 0.059 & 0.072 & 0.065 & 0.205 & 9.170 & 45.73 & $-$1.55 & 43.21\\
133806.59$-$012412.8 & 0.451 & 19.05 & $-$23.14 & 0.172 & 0.206 & 0.192 & 0.505 & 0.000 & 0.000 & 0.000 & 0.000 & 8.890 & 45.37 & $-$1.63 &100.00\\
134233.70$-$001148.0 & 0.516 & 19.13 & $-$23.24 & 0.070 & 0.086 & 0.072 & 0.364 & 0.078 & 0.100 & 0.086 & 0.342 & 8.710 & 45.38 & $-$1.44 & 28.95\\
134318.45$-$005933.6 & 0.697 & 18.75 & $-$24.26 & 0.173 & 0.182 & 0.140 & 0.550 & 0.145 & 0.172 & 0.138 & 0.581 & 8.290 & 45.57 & $-$0.83 &100.00\\
134354.19$-$012759.1 & 0.731 & 18.60 & $-$24.46 & 0.052 & 0.063 & 0.052 & 0.231 & 0.000 & 0.000 & 0.000 & 0.000 & 8.210 & 45.71 & $-$0.61 & 81.53\\
134818.25$-$002441.9 & 0.738 & 18.62 & $-$24.54 & 0.068 & 0.079 & 0.062 & 0.290 & 0.072 & 0.088 & 0.075 & 0.353 & 8.140 & 45.73 & $-$0.52 & 71.63\\
135028.88$-$012958.1 & 0.677 & 19.25 & $-$23.75 & 0.103 & 0.131 & 0.107 & 0.517 & 0.127 & 0.145 & 0.132 & 0.498 & 8.760 & 45.59 & $-$1.28 &100.00\\
135727.37$-$012639.9 & 0.147 & 17.85 & $-$21.81 & 0.035 & 0.045 & 0.040 & 0.157 & 0.061 & 0.074 & 0.060 & 0.261 & 7.300 & 44.67 & $-$0.74 & 98.91\\
135830.46$-$012255.7 & 0.658 & 18.95 & $-$24.00 & 0.070 & 0.084 & 0.073 & 0.273 & 0.000 & 0.000 & 0.000 & 0.000 & 8.120 & 45.51 & $-$0.72 & 57.49\\
140025.53$-$012957.0 & 0.584 & 18.46 & $-$24.30 & 0.044 & 0.052 & 0.045 & 0.159 & 0.045 & 0.056 & 0.047 & 0.226 & 8.400 & 45.73 & $-$0.78 & 69.77\\
140252.06$-$015258.1 & 0.402 & 19.14 & $-$22.69 & 0.079 & 0.099 & 0.087 & 0.334 & 0.000 & 0.000 & 0.000 & 0.000 & 8.500 & 45.14 & $-$1.47 & 99.65\\
142526.50$-$004421.5 & 0.567 & 19.14 & $-$23.64 & 0.061 & 0.071 & 0.065 & 0.253 & 0.000 & 0.000 & 0.000 & 0.000 & 8.100 & 45.58 & $-$0.63 & 76.31\\
140614.86$-$012937.7 & 0.552 & 18.52 & $-$24.06 & 0.059 & 0.074 & 0.066 & 0.237 & 0.067 & 0.082 & 0.075 & 0.266 & 8.600 & 45.70 & $-$1.01 & 99.98\\
\hline
\end{tabular} \\
\end{minipage}
\end{table}
\end{landscape}

\newpage

\begin{landscape}
\begin{table}
\begin{minipage}{20cm}
\contcaption{}
\begin{tabular}{lllllllllllllllr}
\hline
SloanID (SDSS J) & $z$ & $m_{r}$ & $M_{i}$ & $\sigma_{\Delta R}$ & $\left<\Delta R\right>$ & Med$(\Delta R)$ & Max$(\Delta R)$ &
$\sigma_{\Delta V}$ & $\left<\Delta V\right>$ & Med$(\Delta V)$ & Max$(\Delta V)$ & 
$\log M_{\rm BH}$ & $\log L_{\rm bol}$ & $\log \eta$ & GCL \\
 (1) & (2) & (3) & (4) & (5) & (6) & (7) & (8) & (9) & (10) & (11) & (12) & (13) & (14) & (15) & (16) \\ 
\hline
140923.51$-$012430.5 & 0.405 & 19.14 & $-$22.71 & 0.158 & 0.155 & 0.108 & 0.476 & 0.000 & 0.000 & 0.000 & 0.000 & 8.870 & 45.24 & $-$1.74 &100.00\\
141238.77$-$012033.9 & 0.139 & 18.25 & $-$21.37 & 0.053 & 0.065 & 0.055 & 0.196 & 0.069 & 0.081 & 0.069 & 0.242 & 8.110 & 44.48 & $-$1.74 & 95.42\\
141638.18$-$005352.9 & 0.552 & 18.71 & $-$23.78 & 0.179 & 0.191 & 0.140 & 0.604 & 0.197 & 0.235 & 0.226 & 0.686 & 8.400 & 45.47 & $-$1.04 &100.00\\
141647.59$-$012657.4 & 0.376 & 19.30 & $-$22.36 & 0.113 & 0.151 & 0.117 & 0.489 & 0.000 & 0.000 & 0.000 & 0.000 & 8.780 & 45.08 & $-$1.81 & 99.99\\
141949.83$-$000643.7 & 0.515 & 18.91 & $-$23.48 & 0.137 & 0.177 & 0.102 & 0.614 & 0.108 & 0.143 & 0.087 & 0.602 & 9.290 & 45.61 & $-$1.79 &100.00\\
142352.01$-$000705.2 & 0.680 & 18.89 & $-$23.90 & 0.089 & 0.112 & 0.090 & 0.388 & 0.099 & 0.118 & 0.095 & 0.421 & 8.330 & 45.55 & $-$0.89 & 84.43\\
142441.20$-$000727.1 & 0.318 & 18.13 & $-$22.86 & 0.050 & 0.060 & 0.052 & 0.242 & 0.072 & 0.086 & 0.072 & 0.297 & 7.430 & 45.22 & $-$0.32 & 76.20\\
142726.64$-$001950.0   & 0.222 & 18.86 & $-$21.85 & 0.184 & 0.217 & 0.184 & 0.569 & 0.000 & 0.000 & 0.000 & 0.000 & 7.560 & 44.52 & $-$1.15 &100.00\\
143233.96$-$012145.4 & 0.521 & 18.95 & $-$23.65 & 0.114 & 0.125 & 0.109 & 0.367 & 0.140 & 0.159 & 0.141 & 0.467 & 9.000 & 45.67 & $-$1.44 &100.00\\
143306.57$-$012922.9 & 0.627 & 19.09 & $-$23.66 & 0.111 & 0.133 & 0.115 & 0.453 & 0.114 & 0.129 & 0.112 & 0.417 & 8.090 & 45.55 & $-$0.65 &100.00\\
145704.63$-$012253.0 & 0.525 & 18.67 & $-$23.70 & 0.122 & 0.158 & 0.136 & 0.551 & 0.139 & 0.172 & 0.140 & 0.607 & 8.410 & 45.50 & $-$1.02 &100.00\\
145857.84$-$013419.3 & 0.276 & 18.59 & $-$22.23 & 0.044 & 0.053 & 0.048 & 0.167 & 0.081 & 0.101 & 0.084 & 0.379 & 7.590 & 44.99 & $-$0.71 & 86.43\\
145859.95$-$001504.3 & 0.481 & 18.73 & $-$23.56 & 0.106 & 0.135 & 0.107 & 0.394 & 0.157 & 0.193 & 0.152 & 0.555 & 8.730 & 45.36 & $-$1.48 & 99.90\\
145925.04$-$015650.0 & 0.539 & 18.76 & $-$23.65 & 0.098 & 0.118 & 0.095 & 0.600 & 0.000 & 0.000 & 0.000 & 0.000 & 7.960 & 45.56 & $-$0.51 & 99.99\\
151538.94$-$001240.8 & 0.435 & 19.15 & $-$22.97 & 0.053 & 0.065 & 0.055 & 0.234 & 0.000 & 0.000 & 0.000 & 0.000 & 8.890 & 45.42 & $-$1.58 & 58.49\\
151750.63$-$015534.6 & 0.554 & 19.01 & $-$23.68 & 0.103 & 0.128 & 0.110 & 0.361 & 0.000 & 0.000 & 0.000 & 0.000 & 7.950 & 45.48 & $-$0.58 &100.00\\
152018.65$-$015037.4 & 0.632 & 18.99 & $-$23.92 & 0.078 & 0.089 & 0.083 & 0.283 & 0.000 & 0.000 & 0.000 & 0.000 & 7.900 & 45.57 & $-$0.44 & 72.81\\
152035.34$-$002040.1 & 0.130 & 17.17 & $-$22.33 & 0.051 & 0.064 & 0.052 & 0.249 & 0.041 & 0.048 & 0.037 & 0.226 & 7.990 & 44.80 & $-$1.30 & 83.43\\
152809.55$-$000044.9 & 0.605 & 18.07 & $-$24.82 & 0.068 & 0.081 & 0.074 & 0.237 & 0.085 & 0.098 & 0.088 & 0.321 & 8.360 & 45.82 & $-$0.65 &100.00\\
152942.75$-$015934.0 & 0.270 & 19.06 & $-$21.84 & 0.073 & 0.087 & 0.079 & 0.265 & 0.000 & 0.000 & 0.000 & 0.000 & 7.680 & 44.87 & $-$0.92 & 96.63\\
\hline
\end{tabular}\\
\end{minipage}
\end{table}
\end{landscape}

\section{Analysis}
\label{section:analysis}

\subsection{Variability--black hole mass correlation}

\begin{figure}
\begin{center}
\includegraphics[width=8.4 truecm]{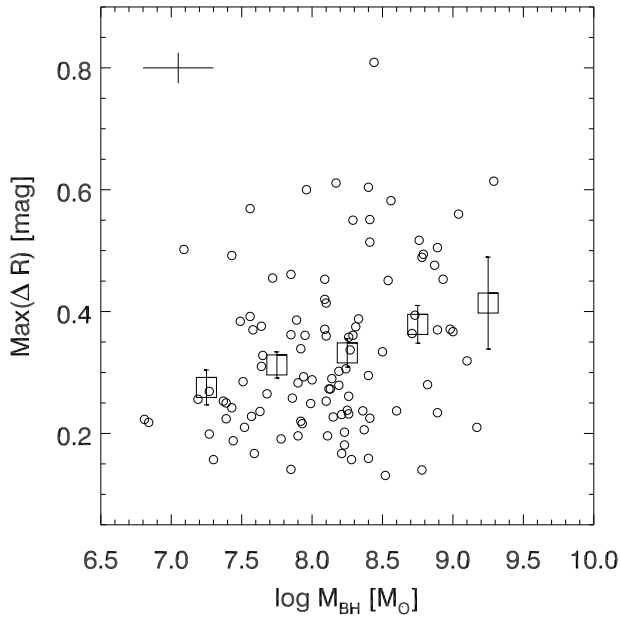}
\end{center}
\caption{The maximum variability amplitude in the $R$-filter as a
function of black hole mass.  Overplotted with open squares are the
mean Max$(\Delta R)$ in five black hole mass bins with the error bars
signifying the error in the mean. The cross in the
upper left-hand corner indicates a typical error bar for the open circles.}
\label{figure:fig5}
\end{figure}

In order to investigate whether quasar variability correlates with any
of the AGN parameters black hole mass, bolometric luminosity or the
Eddington ratio, we calculated Pearson correlation coefficients.
First of all, we note that there exists a number of strong but
uninteresting correlations in the complete correlation matrix that
reflect known selection effects in flux-limited quasar samples (e.g.,
the increase in luminosity with redshift). Secondly, we also measure
significant correlations between quasar variability amplitude and
black hole mass.  The correlation coefficients for the most
interesting parameters are shown in Table~\ref{table:t2}. The
strongest correlation is that between variability as measured in terms
of Max$(\Delta R)$ (i.e.\ maximum variability amplitude) and black
hole mass. The correlation coefficient is 0.285, which corresponds to
a Student's t-statistic of 3.0 and a two-sided probability of arising
by chance of 0.3\%. The correlation between variability amplitude and
black hole mass is therefore present at the $\approx 3\sigma$
level. Other measures of variability have qualitatively similar, but
somewhat less significant, correlations with black hole mass.
Fig.~\ref{figure:fig5} shows the maximum variability
amplitude as a function of black hole mass. There is not a clear
linear relationship, but it is evident that the sources displaying the
most variability have, on average, higher black hole masses.
We also evaluated the correlation matrix using the variability
parameters as determined in the $V$-filter, and obtain very similar
results to the $R$-filter.

\begin{figure*}
\begin{center}
\includegraphics[width=17.5 truecm]{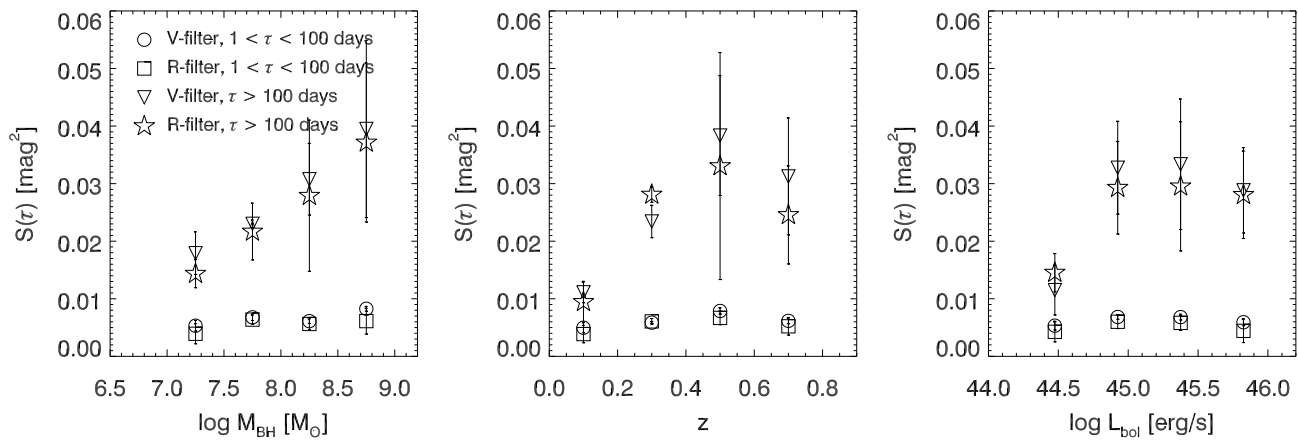}
\end{center}
\caption{Structure functions in $V$ and $R$ 
as a function of black hole mass (left), redshift (centre) and bolometric
luminosity (right).  The increase in variability is clearly seen for
higher black hole masses at longer time lags ($\tau > 100$ days).}
\label{figure:fig6}
\end{figure*}

Further support for the correlation between variability amplitude and
black hole mass is provided by additional analysis of the structure
function.  For each filter, we compute the structure function for two
different bins in rest-frame time lag (1--100 and $>100$ days), with
each time lag bin separated into several smaller bins of black hole
mass.  This allows us to investigate variability as a function of both
rest-frame time lag and black hole mass. We find a significantly
rising trend of variability with black hole mass, see the left-hand
panel of Fig.~\ref{figure:fig6}.  In this figure, it can also
be seen that the main contribution to the variability--black hole mass
correlation comes from the longer time lags at $\tau > 100$ days. At
shorter time lags the correlation is probably diluted by random
measurement noise which is not expected to correlate with AGN
properties.

\subsection{Correlations with redshift and bolometric luminosity}

The middle panel of Fig.~\ref{figure:fig6} shows quasar
variability as a function of redshift, and it is clear that the known
correlation between variability and redshift is present in our sample.

The last panel of Fig.~\ref{figure:fig6} shows variability as a
function of quasar bolometric luminosity, and this figure agrees with
the result from the correlation analysis that there is no significant
correlation present between variability and bolometric luminosity.  We
do however expect an anti-correlation between variability and quasar
luminosity as brought up in the introduction, but the strong
luminosity--redshift correlation in the sample counters it.  This is
supported by a partial Spearman correlation analysis revealing that
Max$(\Delta R)$ and bolometric luminosity at constant redshift are
anti-correlated with a partial Spearman's correlation coefficient of
$-$0.212 and a probability of 0.2\% of the anti-correlation being
caused by underlying correlations with redshift. The
variability--luminosity anti-correlation is therefore detected in our
sample when redshift effects are accounted for.

\begin{table}
 \centering
 \begin{minipage}{80mm}
	\caption{Pearson correlation coefficients between variability
parameters and the AGN parameters black hole mass, bolometric
luminosity and Eddington ratio, $\eta$. Two-sided probabilities, $P$,
that the correlations have arisen by chance are given immediately
under each row of correlation coefficients, and were calculated based
on the Student's t-statistic.}
  \begin{tabular}{@{}lccc@{}}
  \hline
 Variability & $\log M_{\rm BH}$ & $\log L_{\rm bol}$ & $\log \eta$ \\
  \hline

$\sigma_{\Delta R}$ & 0.248  & 0.102 & $-$0.188 \\
$P$                   &  0.011      & 0.303  &  0.057   \\
 $<\Delta R>$        &  0.273      & 0.137 & $-$0.184 \\
$P$                    & 0.005     & 0.164 & 0.062 \\
 Med$(\Delta R)$     &  0.229      & 0.142 & $-$0.129 \\
$P$                    &  0.019      & 0.150  & 0.193 \\
 Max$(\Delta R)$     & {\bf 0.285} & 0.146 & $-$0.189 \\
$P$                    & 0.003 & 0.139 & 0.055 \\
GCL                 & 0.126      & 0.032 & $-$0.114 \\
$P$                      & 0.203 &  0.746 & 0.250 \\
\hline
\end{tabular}
\label{table:t2}
\end{minipage}
\end{table}

\section{Discussion}
\label{section:discussion}

\subsection{Selection effects}

Could the correlation between variability and black hole mass be
caused through more primary correlations with another variable?  For
instance, variability and redshift are known to be correlated, and
Fig.~\ref{figure:fig2} shows that black hole mass is indeed
correlated with redshift in our sample (Spearman's $\rho=0.466$ with
probability of correlation arising by chance 6$\times$10$^{-7}$).  As
discussed in the introduction, the redshift--variability correlation
comes about as a combination of the hardening (bluening) of quasar
spectra in their brighter phases and the fact that a fixed passband in
the observers frame probes progressively shorter rest-frame
wavelengths at higher redshifts. Because higher-redshift quasars in
flux-limited samples are sampled at bluer rest-frame wavelengths where
they are more variable, a variability--redshift correlation emerges.
This could imply that a bias with redshift causes the more massive
black holes to be associated with stronger variability.

In order to check whether the variability--redshift correlation in the
sample is causing the variability--black hole mass correlation, we
evaluate Spearman's partial correlation coefficient between black hole
mass and variability amplitude, at constant redshift.  We obtain
$\rho_{Max(\Delta R) M,z} = 0.203$ with a probability of 0.0035 that
the variability--black hole mass correlation is caused by underlying
correlations with redshift. Hence the variability--black hole mass 
correlation does not appear to be caused by underlying correlations 
with redshift. 

We also considered the possibility that the correlation could arise as
a result of a more primary correlation with luminosity, because black
hole mass is also correlated with bolometric luminosity in our sample
($\rho = 0.513$ with a probability of arising by chance of
2.6$\times$10$^{-8}$).  A partial Spearman correlation analysis shows
that the correlation coefficient between Max$(\Delta R)$ and $M_{\rm
BH}$ at constant $L_{\rm bol}$ is 0.254 with the probability of the
correlation being caused by underlying correlations with $L_{\rm bol}$
being 2$\times$10$^{-4}$.

Additional multivariate regression analysis and principle component
analysis support the conclusions of the partial correlation analysis.
When predicting the Max$(\Delta R)$ parameter based on both black hole
mass and redshift, only black hole mass is significant, still at the
two per cent level.  There is thus no convincing evidence that the
variability--black hole mass correlation is caused by simple selection
effects.

\subsection{Residual Host Galaxy Contamination}

The lowest luminosity AGN in our SDSS-QUEST1 sample are Seyfert
galaxies and their spectra may suffer from some level of host galaxy
contamination despite our efforts to include a host galaxy component
in the spectral fitting.  Host galaxy contamination would occur
preferentially for the lowest luminosity AGNs which are also those
with the largest timescales sampled and tending to have smaller black
hole masses.  Two offsetting effects may occur: the stronger effect,
depending linearly on the host-to-AGN ratio, dilutes the variability;
the weaker effect, depending on the host-to-AGN ratio to the 0.7 power
\citep{kaspi00,kaspi05}, overestimates the black hole mass.

In general we were conservative during the fitting and used zero
initial host galaxy contribution with small steps, and only a few
spectra required much host galaxy contribution.  Furthermore, the
QUEST1 variability catalogue does not include objects regarded as
resolved, and only a few SDSS spectra display any clear host galaxy
features. Finally, the trend between black hole mass and variability
is still present within a sub-sample of higher luminosity quasars
which are expected to be less affected by host galaxy contamination
\citep{vandenberk06}.  Host galaxy contamination has however been
noted to have significant effects on measurements of AGN luminosities
in more luminous PG quasars \citep[e.g.][]{bentz06}, so it is a
potential concern. We conclude that even though host galaxy
contamination could artificially enhance a variability
amplitude--black hole mass correlation, it is a weak effect and
unlikely to be important for our sample.

\subsection{Physical models of quasar variability}

Studies have shown that X-ray, UV and optical quasar flux variations
are approximately simultaneous, supporting a model where reprocessing
is essential. In this model, the optical and UV flux originate in an
accretion disk which is irradiated by an X-ray variable disk
corona. Short-timescale variations in this model are explained as
either flares or variations in the optical depth of the corona.
Longer timescale variations, on the order of months to years, may be
related to the propagation of the shorter-timescale variations
\citep[see][and references therein]{ulrich97}.
In summary, a number of
models for quasar optical variability exists but there are no clear
predictions relating variability amplitude and black hole mass.

There are predictions, however, of how different sources of optical
variability can be associated with different characteristic timescales
and many of these timescales depend on black hole mass. \citet{cp01}
list several (light crossing time, accretion timescale, orbital
timescale and accretion disk thermal timescale, see also
\citet{krolik98}), and attempt to define a relationship between black
hole mass and characteristic variability timescale. Studying a sample
of 10 well-monitored AGN, they report evidence of characteristic
optical variability timescales correlating with black hole mass.  The
timescales they are studying range from a few weeks for AGN with black
hole masses of $\sim10^7 M_{\odot}$ to a few months for AGN with black
hole masses of $\sim10^8 M_{\odot}$, and are roughly consistent with
accretion disk thermal timescales.

On timescales of a few weeks, the QUEST1 light curves are dominated by
observational uncertainties, and timescales of several months are not
well sampled.  The QUEST1 time sampling is far from complete,
essentially sampling timescales of weeks and 1--2 years, with little
in between and nothing longer. The timescales that are well sampled,
and which possess the most variability, are 1--2 years 
\citep[also discussed by][]{rengstorf06}.  AGN with
black hole masses of $\sim10^8 M_{\odot}$ to $\sim10^9 M_{\odot}$,
which we have in our sample, would be expected to have accretion disk
thermal characteristic timescales on the order of 1--2 years.  If the
\citet{cp01} result is correct, we may be seeing a correlation
resulting from the time sampling of QUEST1. If we are biased toward
detecting intrinsic quasar variability only at longer timescales,
i.e.\ $> 100$ days, and if there is a relation between characteristic
variability timescale and black hole mass, we may be preferentially
detecting strong variability in the AGN with higher black hole masses.
The correlation we observe could therefore be a manifestation of the black
hole mass dependence on AGN physical timescales.

There are three arguments against this:
1) The relation between characteristic variability timescales and black 
hole mass is highly speculative and not based on anything that has been 
well established yet.
2) The variability--black hole mass correlation 
is present also in the part of the sample having $M_{\rm BH} > 10^{8}$ 
M$_{\odot}$, although with lower significance than for the complete sample.
Hence the correlation is not formed solely by the presence of quasars at 
$M_{\rm BH} < 10^{8}$ M$_{\odot}$ with variability amplitudes comparable 
to the noise level. 3) 
It is actually the lower redshift quasars (biased toward 
lower black hole masses) that have the best sampling of rest frame 
time lags over the time base line of the QUEST1 survey. 
The lower redshift quasars should therefore make a larger 
contribution to the structure function at intermediate to 
longer time lags as compared to the higher redshift ones. This effect would
work in the opposite direction to what is expected due to the incomplete
temporal sampling discussed above. 

We therefore favour an explanation where the 
correlation we observe 
is due to a real relation between variability amplitude and black hole mass. 
However, as there are no models linking variability amplitude and black 
hole mass, 
it is difficult to explore this in more details at this stage.
In order to confirm the validity of a variability--black hole mass
correlation, larger samples of quasars with better and more complete 
temporal sampling are needed.

Our ensemble structure functions do not show evidence for a
characteristic timescale, although data at longer timescales are
needed to confirm a turn-over or flattening in the structure
function. As discussed in Section~\ref{section:sf}, the dips in the
structure function at $\approx400$ and $700$ days is caused by
incomplete sampling of time lags at the end of the 1 and 2 year
periods in the QUEST1 variability survey. The structure function at
the end of each observing period will be biased toward lower-redshift
quasars for which we can probe the longest rest frame time lags.  As we are
probably biased toward the lower-mass black holes at lower redshifts,
the variability--black hole mass correlation could explain the drop in
the structure function at longer time lags.

We also note that the variability--black hole mass correlation may
help explain the result that radio-loud quasars are marginally more
variable than radio-quiets. This follows if the most radio-loud
objects selected from the top of the radio luminosity function are the
ones with the more massive black holes \citep[e.g.][]{lacy01}. 

\subsection{Palomar-Green (PG) Quasars}

We note that good light curves and black hole masses of PG quasars are
available in the literature.  We were able to assemble a sample of 28
PG quasars with variability parameters from \citet{giveon99} and black
hole masses from \citet{vp06}, and looked for corroboration of our
SDSS-QUEST1 results.  We found no significant correlations between
variability and black hole mass for the PG quasars, possibly because
of the smaller sample size.  We did not add these data points to our
SDSS-QUEST1 sample due to differences in parameter space and time
sampling.

\section{Conclusions}
\label{section:conclusions}

In this study we have matched a sample of broad-lined AGN at redshifts
$z<0.75$ from the SDSS \citep{abazajian04} with sources in the 200k
Light Curve Catalogue of the QUEST1 Variability Survey
\citep{rengstorf04b}, yielding a total sample of 104 quasars.  The
Sloan spectra are used to estimate black hole masses from H$\beta$
linewidths and continuum luminosities at rest frame 5100 {\AA}
\citep{vp06}. The light curves are used to evaluate the variability of
each quasar in the sample, as well as the time-dependence on
variability for the whole ensemble of quasars.

We conclude that there is evidence for a correlation between the black
hole mass of a quasar and its variability properties. In particular,
the variability amplitude tends to be larger with increasing black
hole mass, a trend that is most pronounced for the larger timescales
probed by the QUEST1 Variability Survey. The variability--black hole
mass correlation does not appear to be caused by obvious selection
effects, host galaxy contamination or correlations between variability
and luminosity/redshift. 
We have also discussed whether the correlation may be a
manifestation of a relation between black hole mass and accretion disk
thermal time scale. This can come about if the temporal sampling in the 
QUEST1 survey gives rise to a bias toward variability at longer time
scales. There are however no convincing and established arguments supporting 
this explanation, hence we favour a scenario in which the more massive
black holes have larger variability amplitudes. 

The general robustness of the variability--black hole mass correlation
should be confirmed with samples of similar, or larger, size
consisting of objects possessing longer and better temporal sampling
of the light curves. Improvements in technology and a growing
recognition of the untapped power of time domain analysis promise to
provide such data sets in the near future.

\section*{Acknowledgments}

MSB thanks the ESO Scientific Visitor Programme for support and
hospitality, and the National Science Foundation for support through
grant AST-0507781.  The authours also thank the referee for comments
which helped improve the presentation of this work.
This research has made use of the NASA/IPAC
Extragalactic Database (NED) which is operated by the Jet Propulsion
Laboratory, California Institute of Technology, under contract with
the National Aeronautics and Space Administration. We would like to
thank those involved in making the Sloan Digital Sky Survey and the
QUEST1 Variability Survey realities with public databases.

Funding for the SDSS and SDSS-II has been provided by the Alfred
P. Sloan Foundation, the Participating Institutions, the National
Science Foundation, the U.S. Department of Energy, the National
Aeronautics and Space Administration, the Japanese Monbukagakusho, the
Max Planck Society, and the Higher Education Funding Council for
England. The SDSS Web Site is http://www.sdss.org/.  

\bibliographystyle{mn2e} 


\begin{thebibliography}{}

\bibitem[\protect\citeauthoryear{{Abazajian et al.}}{{Abazajian et
  al.}}{2004}]{abazajian04}
{Abazajian} K.,  {et al.} 2004, \aj, 128, 502

\bibitem[\protect\citeauthoryear{{Angione} \& {Smith}}{{Angione} \&
  {Smith}}{1972}]{as72}
{Angione} R.~J.,  {Smith} H.~J.,  1972, in {Evans} D.~S.,  {Wills} D.,
  {Wills} B.~J.,  eds, IAU Symp. 44: External Galaxies and Quasi-Stellar
  Objects {Optical Variability of Twenty-Two Quasi-Stellar Objects}.
pp 171--+

\bibitem[\protect\citeauthoryear{{Aretxaga}, {Cid Fernandes} \&
  {Terlevich}}{{Aretxaga} et~al.}{1997}]{act97}
{Aretxaga} I.,  {Cid Fernandes} R.,    {Terlevich} R.~J.,  1997, \mnras, 286,
  271

\bibitem[\protect\citeauthoryear{{Aretxaga} \& {Terlevich}}{{Aretxaga} \&
  {Terlevich}}{1994}]{at94}
{Aretxaga} I.,  {Terlevich} R.,  1994, \mnras, 269, 462

\bibitem[\protect\citeauthoryear{{Bentz}, {Peterson}, {Pogge}, {Vestergaard} \&
  {Onken}}{{Bentz} et~al.}{2006}]{bentz06}
{Bentz} M.~C.,  {Peterson} B.~M.,  {Pogge} R.~W.,  {Vestergaard} M.,    {Onken}
  C.~A.,  2006, \apj, 644, 133

\bibitem[\protect\citeauthoryear{{Blandford} \& {McKee}}{{Blandford} \&
  {McKee}}{1982}]{bmk82}
{Blandford} R.~D.,  {McKee} C.~F.,  1982, \apj, 255, 419

\bibitem[\protect\citeauthoryear{{Boroson} \& {Green}}{{Boroson} \&
  {Green}}{1992}]{bg92}
{Boroson} T.~A.,  {Green} R.~F.,  1992, \apjs, 80, 109

\bibitem[\protect\citeauthoryear{{Bruzual} \& {Charlot}}{{Bruzual} \&
  {Charlot}}{2003}]{bc03}
{Bruzual} G.,  {Charlot} S.,  2003, \mnras, 344, 1000

\bibitem[\protect\citeauthoryear{{Cid Fernandes}, {Sodr{\'e}} Jr. \& {Vieira da
  Silva} Jr.}{{Cid Fernandes} et~al.}{2000}]{cf00}
{Cid Fernandes} R.,  {Sodr{\'e}} Jr. L.,    {Vieira da Silva} Jr. L.,  2000,
  \apj, 544, 123

\bibitem[\protect\citeauthoryear{{Cid Fernandes}, {Aretxaga} \&
  {Terlevich}}{{Cid Fernandes} et~al.}{1996}]{cf96}
{Cid Fernandes} R.~J.,  {Aretxaga} I.,    {Terlevich} R.,  1996, \mnras, 282,
  1191

\bibitem[\protect\citeauthoryear{{Collier} \& {Peterson}}{{Collier} \&
  {Peterson}}{2001}]{cp01}
{Collier} S.,  {Peterson} B.~M.,  2001, \apj, 555, 775

\bibitem[\protect\citeauthoryear{{Cristiani}, {Trentini}, {La Franca},
  {Aretxaga}, {Andreani}, {Vio} \& {Gemmo}}{{Cristiani}
  et~al.}{1996}]{cristiani96}
{Cristiani} S.,  {Trentini} S.,  {La Franca} F.,  {Aretxaga} I.,  {Andreani}
  P.,  {Vio} R.,    {Gemmo} A.,  1996, \aap, 306, 395

\bibitem[\protect\citeauthoryear{{Cristiani}, {Vio} \& {Andreani}}{{Cristiani}
  et~al.}{1990}]{cristiani90}
{Cristiani} S.,  {Vio} R.,    {Andreani} P.,  1990, \aj, 100, 56

\bibitem[\protect\citeauthoryear{{de Vries}, {Becker}, {White} \& {Loomis}}{{de
  Vries} et~al.}{2005}]{devries05}
{de Vries} W.~H.,  {Becker} R.~H.,  {White} R.~L.,    {Loomis} C.,  2005, \aj,
  129, 615

\bibitem[\protect\citeauthoryear{{Enya}, {Yoshii}, {Kobayashi}, {Minezaki},
  {Suganuma}, {Tomita} \& {Peterson}}{{Enya} et~al.}{2002}]{enya02}
{Enya} K.,  {Yoshii} Y.,  {Kobayashi} Y.,  {Minezaki} T.,  {Suganuma} M.,
  {Tomita} H.,    {Peterson} B.~A.,  2002, \apjs, 141, 45

\bibitem[\protect\citeauthoryear{{Giallongo}, {Trevese} \&
  {Vagnetti}}{{Giallongo} et~al.}{1991}]{giallongo91}
{Giallongo} E.,  {Trevese} D.,    {Vagnetti} F.,  1991, \apj, 377, 345

\bibitem[\protect\citeauthoryear{{Giveon}, {Maoz}, {Kaspi}, {Netzer} \&
  {Smith}}{{Giveon} et~al.}{1999}]{giveon99}
{Giveon} U.,  {Maoz} D.,  {Kaspi} S.,  {Netzer} H.,    {Smith} P.~S.,  1999,
  \mnras, 306, 637

\bibitem[\protect\citeauthoryear{{Helfand}, {Stone}, {Willman}, {White},
  {Becker}, {Price}, {Gregg} \& {McMahon}}{{Helfand} et~al.}{2001}]{helfand01}
{Helfand} D.~J.,  {Stone} R.~P.~S.,  {Willman} B.,  {White} R.~L.,  {Becker}
  R.~H.,  {Price} T.,  {Gregg} M.~D.,    {McMahon} R.~G.,  2001, \aj, 121, 1872

\bibitem[\protect\citeauthoryear{{Hook}, {McMahon}, {Boyle} \& {Irwin}}{{Hook}
  et~al.}{1994}]{hook94}
{Hook} I.~M.,  {McMahon} R.~G.,  {Boyle} B.~J.,    {Irwin} M.~J.,  1994,
  \mnras, 268, 305

\bibitem[\protect\citeauthoryear{{Hughes}, {Aller} \& {Aller}}{{Hughes}
  et~al.}{1992}]{hughes92}
{Hughes} P.~A.,  {Aller} H.~D.,    {Aller} M.~F.,  1992, \apj, 396, 469

\bibitem[\protect\citeauthoryear{{Kaspi}, {Maoz}, {Netzer}, {Peterson},
  {Vestergaard} \& {Jannuzi}}{{Kaspi} et~al.}{2005}]{kaspi05}
{Kaspi} S.,  {Maoz} D.,  {Netzer} H.,  {Peterson} B.~M.,  {Vestergaard} M.,
  {Jannuzi} B.~T.,  2005, \apj, 629, 61

\bibitem[\protect\citeauthoryear{{Kaspi}, {Smith}, {Netzer}, {Maoz}, {Jannuzi}
  \& {Giveon}}{{Kaspi} et~al.}{2000}]{kaspi00}
{Kaspi} S.,  {Smith} P.~S.,  {Netzer} H.,  {Maoz} D.,  {Jannuzi} B.~T.,
  {Giveon} U.,  2000, \apj, 533, 631

\bibitem[\protect\citeauthoryear{{Kawaguchi}, {Mineshige}, {Umemura} \&
  {Turner}}{{Kawaguchi} et~al.}{1998}]{kawaguchi98}
{Kawaguchi} T.,  {Mineshige} S.,  {Umemura} M.,    {Turner} E.~L.,  1998, \apj,
  504, 671

\bibitem[\protect\citeauthoryear{{Kriss}}{{Kriss}}{1994}]{kriss94}
{Kriss} G.,  1994, in {Crabtree} D.~R.,  {Hanisch} R.~J.,   {Barnes} J.,  eds,
  ASP Conf. Ser. 61: Astronomical Data Analysis Software and Systems III
  {Fitting Models to UV and Optical Spectral Data}.
pp 437--+

\bibitem[\protect\citeauthoryear{{Krolik}}{{Krolik}}{1998}]{krolik98}
{Krolik} J.~H.,  1998, {Active Galactic Nuclei: From the Central Black Hole to
  the Galactic Environment}.
Princeton University Press, 1998.

\bibitem[\protect\citeauthoryear{{Lacy}, {Laurent-Muehleisen}, {Ridgway},
  {Becker} \& {White}}{{Lacy} et~al.}{2001}]{lacy01}
{Lacy} M.,  {Laurent-Muehleisen} S.~A.,  {Ridgway} S.~E.,  {Becker} R.~H.,
  {White} R.~L.,  2001, \apjl, 551, L17

\bibitem[\protect\citeauthoryear{{Morgan}, {Kochanek}, {Morgan} \&
  {Falco}}{{Morgan} et~al.}{2006}]{morgan06}
{Morgan} C.~W.,  {Kochanek} C.~S.,  {Morgan} N.~D.,    {Falco} E.~E.,  2006,
  \apj, 647, 874

\bibitem[\protect\citeauthoryear{{Netzer} \& {Peterson}}{{Netzer} \&
  {Peterson}}{1997}]{np97}
{Netzer} H.,  {Peterson} B.~M.,  1997, in {Maoz} D.,  {Sternberg} A.,
  {Leibowitz} E.~M.,  eds, ASSL Vol. 218: Astronomical Time Series
  {Reverberation Mapping and the Physics of Active Galactic Nuclei}.
pp 85--+

\bibitem[\protect\citeauthoryear{{Pereyra}, {Vanden Berk}, {Turnshek},
  {Hillier}, {Wilhite}, {Kron}, {Schneider} \& {Brinkmann}}{{Pereyra}
  et~al.}{2006}]{pereyra06}
{Pereyra} N.~A.,  {Vanden Berk} D.~E.,  {Turnshek} D.~A.,  {Hillier} D.~J.,
  {Wilhite} B.~C.,  {Kron} R.~G.,  {Schneider} D.~P.,    {Brinkmann} J.,  2006,
  \apj, 642, 87

\bibitem[\protect\citeauthoryear{{Peterson}}{{Peterson}}{1997}]{peterson97}
{Peterson} B.~M., 1997, {Book Review: An Introduction to Active Galactic Nuclei}. Cambridge University Press, 1997


\bibitem[\protect\citeauthoryear{{Peterson et al.}}{{Peterson et
  al.}}{2004}]{peterson04}
{Peterson} B.~M.,  {et al.} 2004, \apj, 613, 682

\bibitem[\protect\citeauthoryear{{Refsdal}, {Stabell}, {Pelt} \&
  {Schild}}{{Refsdal} et~al.}{2000}]{refsdal00}
{Refsdal} S.,  {Stabell} R.,  {Pelt} J.,    {Schild} R.,  2000, \aap, 360, 10

\bibitem[\protect\citeauthoryear{{Rengstorf et al.}}{{Rengstorf et
  al.}}{2004a}]{rengstorf04a}
{Rengstorf} A.~W.,  {et al.} 2004a, \apj, 606, 741

\bibitem[\protect\citeauthoryear{{Rengstorf et al.}}{{Rengstorf et
  al.}}{2004b}]{rengstorf04b}
{Rengstorf} A.~W.,  {et al.} 2004b, \apj, 617, 184

\bibitem[\protect\citeauthoryear{{Rengstorf}, {Brunner} \& {Wilhite}}
{{Rengstorf et~al.}}{2006}]{rengstorf06}
{Rengstorf} A.~W.,  {Brunner} R.~J., {Wilhite} B.~C., 2006, \aj, 131, 1923

\bibitem[\protect\citeauthoryear{{Schlegel}, {Finkbeiner} \&
  {Davis}}{{Schlegel} et~al.}{1998}]{sfd98}
{Schlegel} D.~J.,  {Finkbeiner} D.~P.,    {Davis} M.,  1998, \apj, 500, 525

\bibitem[\protect\citeauthoryear{{Schneider et al.}}{{Schneider et
  al.}}{2005}]{schneider05}
{Schneider} D.~P.,  {et al.} 2005, \aj, 130, 367

\bibitem[\protect\citeauthoryear{{Shang et al.}}{{Shang et
  al.}}{2005}]{shang05}
{Shang} Z.,  {et al.} 2005, \apj, 619, 41

\bibitem[\protect\citeauthoryear{{Terlevich}, {Tenorio-Tagle}, {Franco} \&
  {Melnick}}{{Terlevich} et~al.}{1992}]{terlevich92}
{Terlevich} R.,  {Tenorio-Tagle} G.,  {Franco} J.,    {Melnick} J.,  1992,
  \mnras, 255, 713

\bibitem[\protect\citeauthoryear{{Torricelli-Ciamponi}, {Foellmi},
  {Courvoisier} \& {Paltani}}{{Torricelli-Ciamponi}
  et~al.}{2000}]{Torricelli00}
{Torricelli-Ciamponi} G.,  {Foellmi} C.,  {Courvoisier} T.~J.-L.,    {Paltani}
  S.,  2000, \aap, 358, 57

\bibitem[\protect\citeauthoryear{{Tr{\`e}vese} \& {Vagnetti}}{{Tr{\`e}vese} \&
  {Vagnetti}}{2002}]{tv02}
{Tr{\`e}vese} D.,  {Vagnetti} F.,  2002, \apj, 564, 624

\bibitem[\protect\citeauthoryear{{Ulrich}, {Maraschi} \& {Urry}}{{Ulrich}
  et~al.}{1997}]{ulrich97}
{Ulrich} M.-H.,  {Maraschi} L.,    {Urry} C.~M.,  1997, \araa, 35, 445

\bibitem[\protect\citeauthoryear{{Vanden Berk et al.}}{{Vanden Berk
  et al.}}{2004}]{vandenberk04}
{Vanden Berk} D.~E.,  {et al.} 2004, \apj, 601, 692

\bibitem[\protect\citeauthoryear{{Vanden Berk et al.}}{{Vanden Berk
  et al.}}{2006}]{vandenberk06}
{Vanden Berk} D.~E.,  {et al.} 2006, \aj, 131, 84

\bibitem[\protect\citeauthoryear{{Vestergaard}}{{Vestergaard}}{2002}]{vesterga%
ard02}
{Vestergaard} M.,  2002, \apj, 571, 733

\bibitem[\protect\citeauthoryear{{Vestergaard} \& {Peterson}}{{Vestergaard} \&
  {Peterson}}{2006}]{vp06}
{Vestergaard} M.,  {Peterson} B.~M.,  2006, \apj, 641, 689

\bibitem[\protect\citeauthoryear{{Webb} \& {Malkan}}{{Webb} \&
  {Malkan}}{2000}]{wm00}
{Webb} W.,  {Malkan} M.,  2000, \apj, 540, 652

\end{thebibliography}

\label{lastpage}

\end{document}